\documentclass[desactivate]{aa}
\usepackage[utf8]{inputenc}
\usepackage[varg]{txfonts}
\usepackage{graphicx}
\usepackage{hyperref}
\hypersetup{ colorlinks, linkcolor=blue, citecolor=blue }
 \newcommand{\be}{\element[][7]{Be}}
\newcommand{\heq}{\element[][4]{He}} 
 
 \newcommand{\carbon}{\element[][12]{C}}
\newcommand{\oxygen}{\element[][16]{O}}
\newcommand{\calpha}{\carbon (\alpha, \gamma) \oxygen}
\newcommand{\ov}{\alpha_{\mathrm{ov}}} 
 \newcommand{\dd}{\mathrm{d}}

\newcommand{\smass}{M_{\odot}} 
 \newcommand{\gradad}{\nabla_{\rm
ad}} \newcommand{\gradrad}{\nabla_{\mathrm{rad}}}
\newcommand{\dov}{d_{\mathrm{ov}}} \newcommand{\brunt}{Brunt-Väisälä }
\newcommand{\reac}{\carbon(\alpha,\gamma)\oxygen}
\title{Effect of nuclear reactions rates and core boundary mixing on the seismology of Red Clump stars}
\author{Anthony Noll\inst{1}, Sarbani Basu\inst{2}, and Saskia Hekker\inst{1,3}}
\titlerunning{Seismology of Red Clump stars}
\institute{Heidelberger Institut für Theoretische Studien,
Schloss-Wolfsbrunnenweg 35, 69118 Heidelberg, Germany\label{1} \and Department
of Astronomy, Yale University, PO Box 208101, New Haven, CT 06520-8101,
USA\label{2} \and Center for Astronomy (ZAH/LSW), Heidelberg University,
Königstuhl 12, 69117 Heidelberg, Germany\label{3}}
\begin{document}

\abstract{Modeling of the stars in the red clump (RC), i.e. core helium burning
stars that have gone through a He-flash, is challenging, due to the
uncertainties associated with the physical processes happening in their core and
during the helium flash. By probing the internal stellar structure,
asteroseismology allows us to constrain the core properties of RC stars and
eventually improve our understanding of this evolutionary phase.}{We aim to
quantify the impact on the seismic properties of the RC stars of the two
main core modeling uncertainties: core boundary mixing, and helium burning
nuclear reaction rates.}{Using the MESA stellar evolution code, we compute
models with different core boundary mixing as well as different $3\alpha$ and
$\reac$ nuclear reaction rates. We investigate the impact of those parameters on
the period spacing $\Delta \Pi$, which is a probe of the region around the
core.}{We find that different core boundary mixing schemes yield significantly
different period spacings, with differences of 30\,s between the maximal $\Delta
\Pi$ value computed with semiconvection and maximal overshoot. We show that
increasing the rate of $\reac$ lengthens the core helium burning phase, which
extends the range of $\Delta \Pi$ covered by the models during their evolution.
This results in a difference of 10\,s between the models computed with a nominal
rate and a rate multiplied by 2, which is larger than the observational
uncertainties. The effect of changing the 3$\alpha$ reaction rate is
comparatively small.}{The core boundary mixing is the main source of uncertainty
regarding the seismic modeling of RC stars. Moreover, the effect of the
$\reac$ is non-negligible, even if difficult to disentangle from the effect of
the mixing. Such degeneracy could be raised in the future, thanks to new seismic
data from the PLATO mission and theoretical constraints from numerical
simulations.}
\maketitle

\section{Introduction}

Stars in the red clump (RC) are somewhat high-metallicity (higher than
approximately $-0.6\,\mathrm{dex}$), low-mass ($\lesssim 1.8\,\smass$) stars
that are in the process of fusing helium in their cores. Their key properties
derive from the fact that their progenitors had a core made of degenerate matter
when they were in the red giant branch (RGB) phase; this property results in
helium core masses that are largely independent of the total mass. Consequently,
properties of the RC stars have little dependency on the stellar mass, which
makes them targets of significant astrophysical importance. RC stars have been
proposed as standard candles (thanks to their narrow range of luminosity) and
reddening indicators \citep{Cannon1970}, as well as tracers of the chemical
evolution of the Galaxy \citep[e.g.,][]{Nidever2014}. Since these stars are also
the progenitors of later stages of stellar evolutions, improving their models is
key to putting better constraints on the structure of, for instance, white dwarf
stars.

However, the modeling of several physical processes occurring within RC stars is
still uncertain. The first main unknown is mixing at the boundary of the
convective core, which we refer to as core boundary mixing (CBM).
\cite{Schwarzschild1969} found the presence of a semiconvective region around
the core, whose properties have been studied by
\cite{Castellani1971,Castellani1971b}. However, the need for semiconvection was
questioned by \cite{Bressan1986}, who showed that including a strong overshoot
(over a distance of the order of a pressure scale height) allows reproducing the
star counts of globular clusters with no semiconvective region. 

Other uncertainties in the modeling of RC stars are nuclear reactions rates,
specifically the rates of the helium-burning 3$\alpha$ and $\reac$ reactions.
The rate of the first reaction is somewhat well determined. However, the rate of
the latter is notoriously ill-defined experimentally because the rates need to
be extrapolated to low-energy stellar regimes, which is made more
difficult by the presence of several interfering resonant states
\citep{Angulo1999,Kunz2002,deBoer2017}. 

Thanks to data from the space missions of the last decade, and especially
\emph{Kepler} \citep{borucki10}, it has been possible to retrieve the seismic
properties of thousands of RC stars. RC stars are solar-like oscillators, with
their modes being stochastically excited by the turbulence in the outer
convective envelope of these stars. Notably, their non-radial modes have a mixed
nature, acting as gravity modes in the core and pressure modes in layers above.
Therefore, the study of these modes allows us to constrain the properties of the
deep layers of the star, and especially probe the CBM. A striking result from
the \emph{Kepler} asteroseismic observables is that a large core extension is
required to reproduce the observed seismic properties
\citep[e.g.,][]{Montalban2013,Constantino2015,Bossini2015}. 

In this paper, we build on the works of \cite{Constantino2015,Constantino2017}
and \cite{Bossini2015,Bossini2017} to investigate the combined effects of core
boundary mixing, nuclear reactions rates and composition on the period spacing
of RC stars. We probe a wide parameter space in order to assess the changes in
the period spacings. We review,  in Sect.~\ref{section_seismic_properties} the
seismic properties of RC stars and what structural features  the period spacing
probe. In Sect.~\ref{section_physics}, after describing the physics of the
models, we explain the properties of the different CBM schemes used and their
effect on the period spacing. In Sect.~\ref{effect_nuclear_reaction}, we study
the effect of varying the rates of $\reac$ and 3$\alpha$ on the period spacing.
In Sect.~\ref{effect_other_parameters}, we look at the effect of the metallicity
on the period spacing. Then, in Sect.~\ref{section_combined}, we look at the
combined effects of CBM, nuclear reaction rates, and metallicity. We end with a
discussion and a conclusion in Sect.~\ref{section_discussion} and
\ref{section_conclusion},  respectively.

\section{Seismic properties of RC stars}
\label{section_seismic_properties}

\begin{figure}
    \centering
    \includegraphics[width=0.45\textwidth]{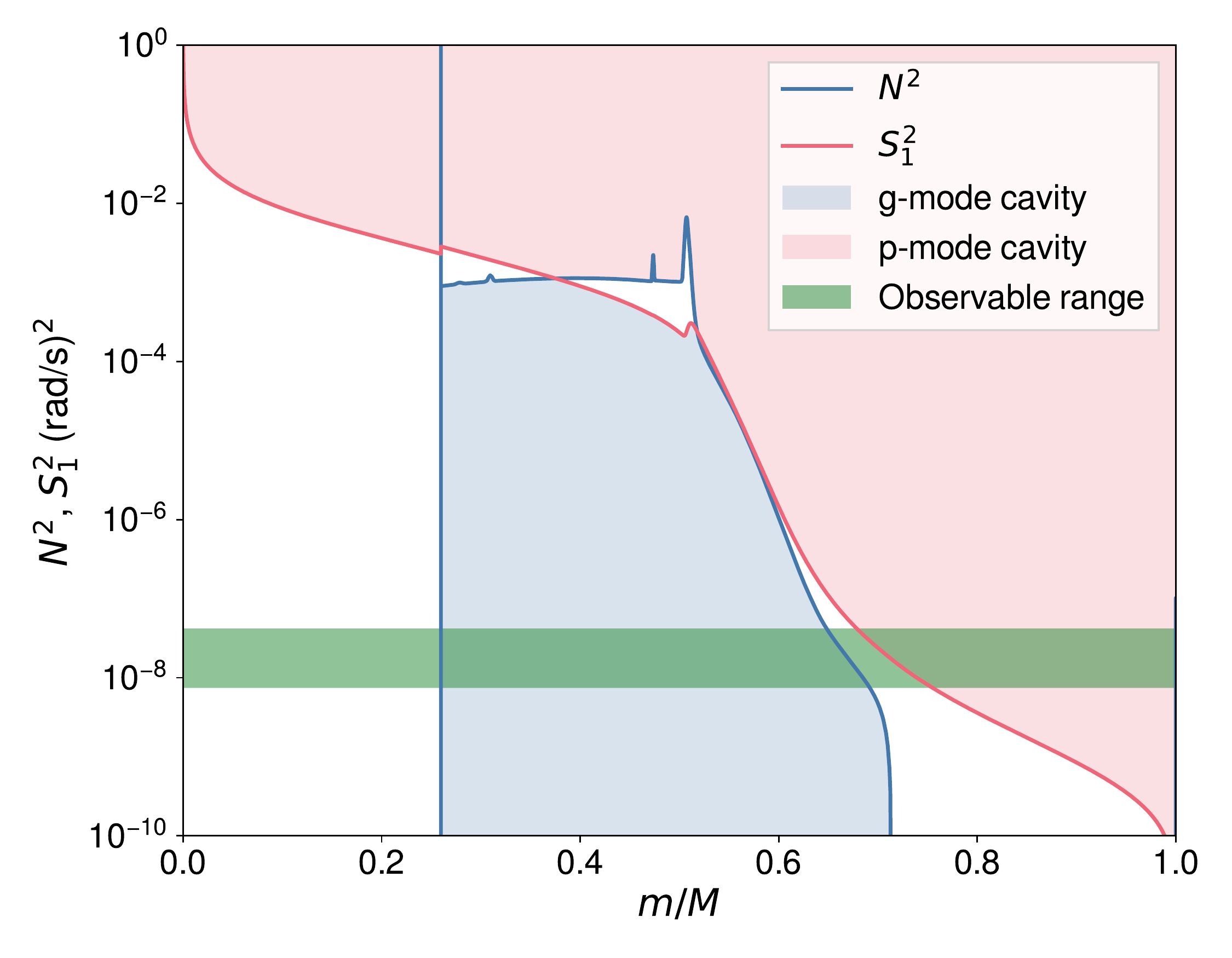}
    \caption{Propagation diagram of an RC star. The \brunt frequency and $\ell=1$ Lamb frequencies are shown with blue and pink lines, respectively. They delimit the g-mode (light blue) and p-mode (light pink) cavities. Finally, the typical range of observed frequencies is shown in green.}
    \label{propagation_diagram}
\end{figure}

Non-radial modes of RC stars are mixed modes, which behave like gravity modes in
the g-mode cavity and like pressure modes in the p-mode cavity. For a mode whose
angular frequency is $\omega$, the g-mode cavity is delimited by $\omega^2 <
N^2, S^2_\ell$, $N$ being the \brunt frequency and $S_\ell$ the Lamb frequency
for a mode with degree $\ell$. Conversely, the p-mode cavity is defined by
$\omega^2 > S^2_\ell, N^2$. In Fig.~\ref{propagation_diagram}, we show those two
cavities in an RC star. The g-mode cavity is located in the radiative region
between the convective core and convective envelope, while the p-mode cavity is
located in the outer layers of the star. The g-modes are therefore more
sensitive to the inner part of the star, while the p-modes rater probe the
envelope properties. 

Pure g-modes are, asymptotically, equally spaced in period \citep{shibahashi79}.
The periods of consecutive modes of degree $\ell$ are separated by a period
spacing $\Delta \Pi_\ell$, which is defined as:
\begin{equation}
    \label{delta_pi_asymp}
    \Delta \Pi_\ell = \frac{2 \pi^2}{\sqrt{\ell (\ell +1 )}} \left( \int_{r_0}^{r_1} \frac{N}{r} \dd r \right)^{-1},
\end{equation}
where $r$ is the radial coordinate and $r_0,r_1$ are the limits of the g-mode
cavity. In the following, we only look at the period spacing of dipole modes,
since we only observe those modes; for simplicity we call this $\Delta \Pi$.
Through the integral of $N/r$ in the g-mode cavity, $\Delta \Pi$ is tightly
related to the size of the convective core. Indeed, a larger convective core
would make the g-mode cavity smaller and accordingly increase $\Delta \Pi$,
making the period spacing a very good probe of the properties near the
convective core and a precise measure of the mass of the helium core
\citep{Montalban2013}. In this paper, we compute $\Delta \Pi$ using
Eq.~(\ref{delta_pi_asymp}). It will be the main seismic observable of interest,
as we study the effect of helium burning reactions which happen in the core of
the star.

Thanks to the \emph{Kepler} satellite data, period spacing can be measured for
thousands of RC stars. It covers a range of period spacing from approximately
230 to 340 seconds, largely independently of the mass \citep{Vrard2016}, as we
can expect from those low-mass stars which had a degenerate core on the red
giant branch.

\section{Physics of the models}
\label{section_physics}

\subsection{General properties}

We computed models  with MESA version r22.11.1
\citep{Paxton2011,Paxton2013,paxton15,paxton18,Paxton2019,Jermyn2023}. The
equation of state is a mix of FreeEOS \citep{Irwin2012} in the envelope and Skye
\citep{Jermyn2021} in the core. The opacities come from the OPAL code
\citep{opal_opacities}, and we use the solar mixture of \cite{gs98}. The
convection regions are computed using the time-dependent local
\cite{Kuhfuss1986} model, adapted in such a way that it reduces to \cite{cox68}
mixing-length prescription at typical stellar evolutionary times. We set the
mixing-length parameter $\alpha_\mathrm{MLT}$ to 1.8. The nuclear reaction rates
are from the \texttt{REACLIB} database\footnote{During this work, it has been
found that the above version of MESA has a faulty implementation of the reverse
3$\alpha$ reaction, which causes a depletion of carbon at the end of the core
helium-burning phase. However, it does not affect us since we do not take this
reaction into account in our nuclear network because its rate is extremely low
at temperatures relevant to our study. We are therefore not concerned by this
issue.} \citep{Cyburt2010}. Any modification in the nuclear reaction rates
described in the text are with respect to these rates.

\subsection{Core boundary mixing}
\begin{figure}
    \centering
    \includegraphics[width=0.45\textwidth]{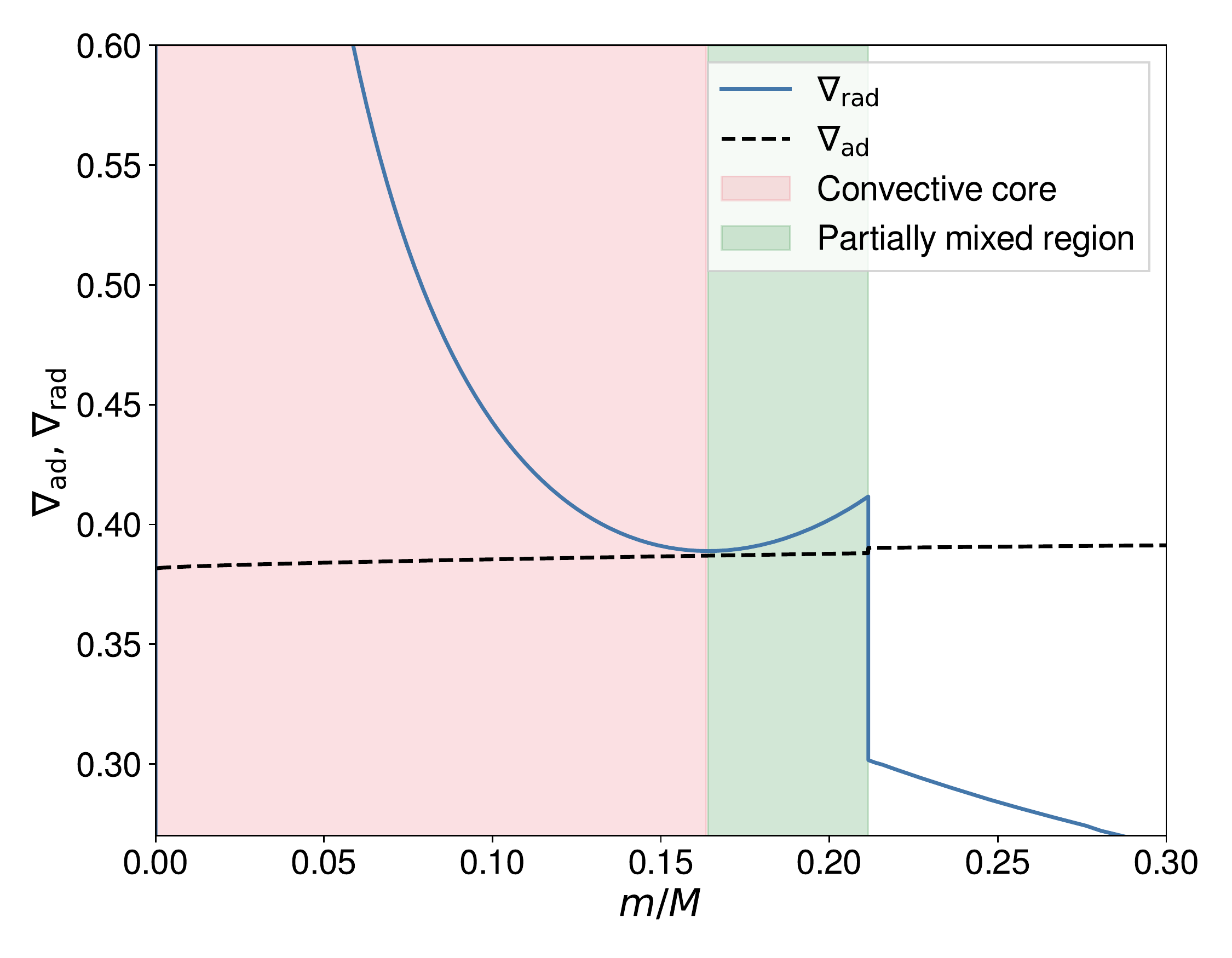}
    \caption{Central structure of a $1\,\smass$ core helium burning star in terms of radiative and adiabatic temperature gradients vs. the fractional mass. The radiative and adiabatic gradients are shown in solid blue and dashed black lines, respectively. We also show the convective core and partially mixed regions in pale pink and green.}
    \label{core_struc_cheb}
\end{figure}

\begin{figure*}
    \centering
    \includegraphics[width=\textwidth]{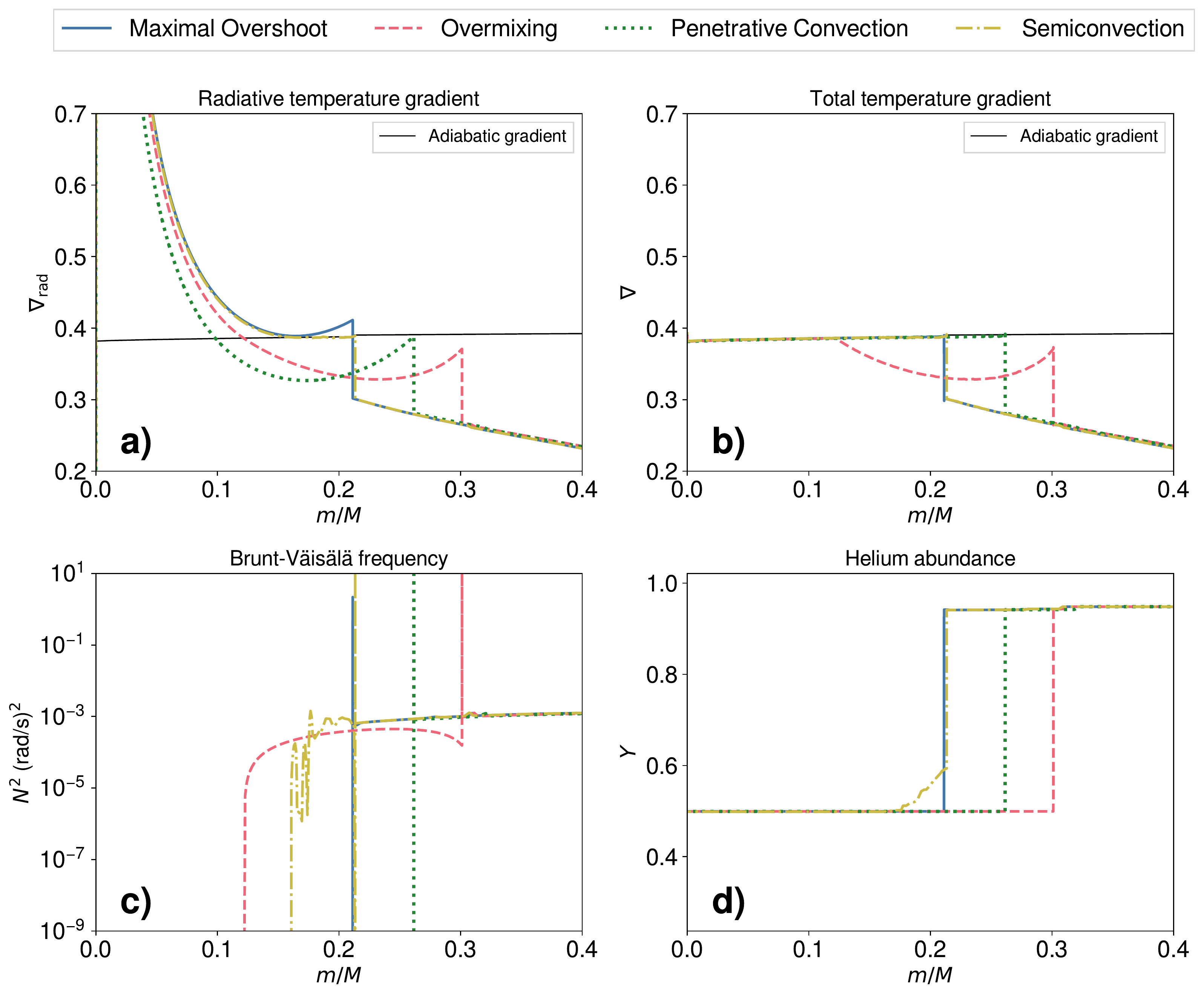}
    \caption{Properties around the core of $1\,\smass$, solar-metallicity models
    computed with the Maximal overshoot (blue, solid),  overmixing (pink,
    dashed),  penetrative convection (green, dotted) and  semiconvection
    (yellow, dash-dotted) core boundary mixing schemes. \emph{Panel a:}
    Radiative and adiabatic gradients. \emph{Panel b:} Temperature gradients
    ($\nabla = \partial \ln T / \partial \ln P$). \emph{Panel c:} \brunt
    frequency profile. \emph{Panel d:} Helium composition profile.}
    \label{mixing_properties}
\end{figure*}
\label{mixing_prescriptions}

During the helium-burning phase, the core becomes enriched in carbon and oxygen,
which increases the opacity and thus the radiative gradient, leading to a
growing core. Once the core reaches a certain size, the radiative gradient
within it is not monotonic anymore and, after a minimum, increases with radius.
Such a radiative temperature gradient is illustrated in
Fig.~\ref{core_struc_cheb}. The core is then split in two regions: the fully
mixed core (between the center and the minimum of $\gradrad$ ) and the
\emph{partially mixed} (PM) region, between the minimum of $\gradrad$ and the
radiative region. The properties of the PM region have a significant effect on
the \brunt frequency profile, and therefore on the asymptotic period spacing.
Thus, we examine the effect of different partial mixing prescriptions in this
work. As an illustration, we show in Fig.~\ref{mixing_properties} the profile of
different properties (temperature gradients, \brunt frequency, and composition)
for the four different mixing prescriptions, which we describe below.

\paragraph*{Semiconvection:} 
\cite{Schwarzschild1969} found that the PM region is a \emph{semiconvective}
region, which is buoyantly neutral (i.e., $\gradrad = \gradad$). To achieve
that, we use in this work the convective premixing (CPM) scheme
\citep{Paxton2019}. This scheme treats every convective boundary with an
iterative approach (within an evolutionary timestep) such that the Schwarzschild
(or Ledoux) criterion is respected on the convective side of the boundary (in
accordance with the recommendation from \citealt{gabriel14}). To explain how the
CPM scheme leads to the formation of a semiconvective region, let us start from
a situation similar to the one shown in Fig.~\ref{core_struc_cheb}. One can see
that, at the outer boundary of the partially PM region, $\gradrad > \gradad$
which causes the boundary to move to a larger radius due to overshooting. This
brings fresh unburned helium in the PM region that lowers the opacity in the PM
region and therefore the radiative gradient. As a consequence, the inner PM
region boundary, which was defined by the local minimum of the radiative
gradient and was marginally unstable, moves to a larger radius, leaving a
buoyantly neutral region in its wake. This process repeats, until the full PM
region is buoyantly neutral. This mechanism of mixing through successive
shrinking convective shells has been described by \cite{Gabriel1970} for massive
star models, and is similar to the \cite{Castellani1985} semiconvection scheme
for core helium burning (CHeB) stars.

Semiconvection is sensitive to the core breathing pulses (CBP), which occur at
the end of the core-helium burning phase. These pulses are sudden increases of
the core size that are caused by the strong increase of energy produced by the
$3\alpha$ triple alpha reaction when a small (in absolute sense) but high (in
relative sense) quantity of helium is injected in the core. This unstable
behavior impacts both $\Delta \Pi$ and the duration of the core-helium burning
phase. However, CBP seem to be ruled out by observations of globular clusters
and asteroseismology \citep{Caputo1989,Cassisi2001,Constantino2017}. For that
reason, we avoid the increase of central helium abundance during the CHeB phase.
This helps to reduce the number of CBP, without totally eliminating them.
Finally, \cite{Constantino2015} observed that mode trapping may occur in the
semiconvective region, which may impact the seismic properties of the star. We
discuss this process in Sect.~\ref{trapping}.

\paragraph*{Maximal overshoot:} This scheme was introduced by
\cite{Constantino2015} to reconcile model predictions with seismic observations,
and results in a large asymptotic period spacing. It is defined by the following
algorithm: once the local $\gradrad$ minimum appears in the core, it extends the
core such that the local minimum of the radiative gradient is equal to the
adiabatic gradient. A core larger than this would decrease the radiative
gradient, which would split the convective region in two and thus lower the
period spacing. This scheme is ad-hoc and quite nonphysical, notably by the fact
that the Schwarzschild criterion is not respected on the convective side of the
external boundary (see Fig.~\ref{mixing_properties} a). However, it leads to a
behavior of the period spacing that is similar to the one resulting from a
potential mode trapping in the semiconvective region, which we discuss in
Sect.~\ref{trapping}.

\paragraph*{Overmixing and penetrative convection:} The core can be extended
over a certain distance $\dov$, defined by a free parameter $\ov$ such that
$\dov = \ov H_p$, with $H_p$ the pressure scale height. The thermal
stratification in the extension region can either be radiative (i.e., $\nabla =
\gradrad$) or adiabatic (i.e., $\nabla = \gradad$). The first case is known as
\emph{overmixing} while the second is known as \emph{penetrative convection}. We
note that, at some point in the evolution, a convective shell appears at the
outer boundary of the overshooting region. This shell becomes a semiconvective
region, when using a convective premixing scheme. Note that the
overmixing and penetrative convection schemes used here are different from the
ones used in \cite{Bossini2015,Bossini2017}, due to the presence of this
semiconvective region. More details on the effect of $\ov$ and the core
boundary scheme are given in Sect.~\ref{section_effect_aov}.

\begin{figure}
    \centering
    \includegraphics[width=0.45\textwidth]{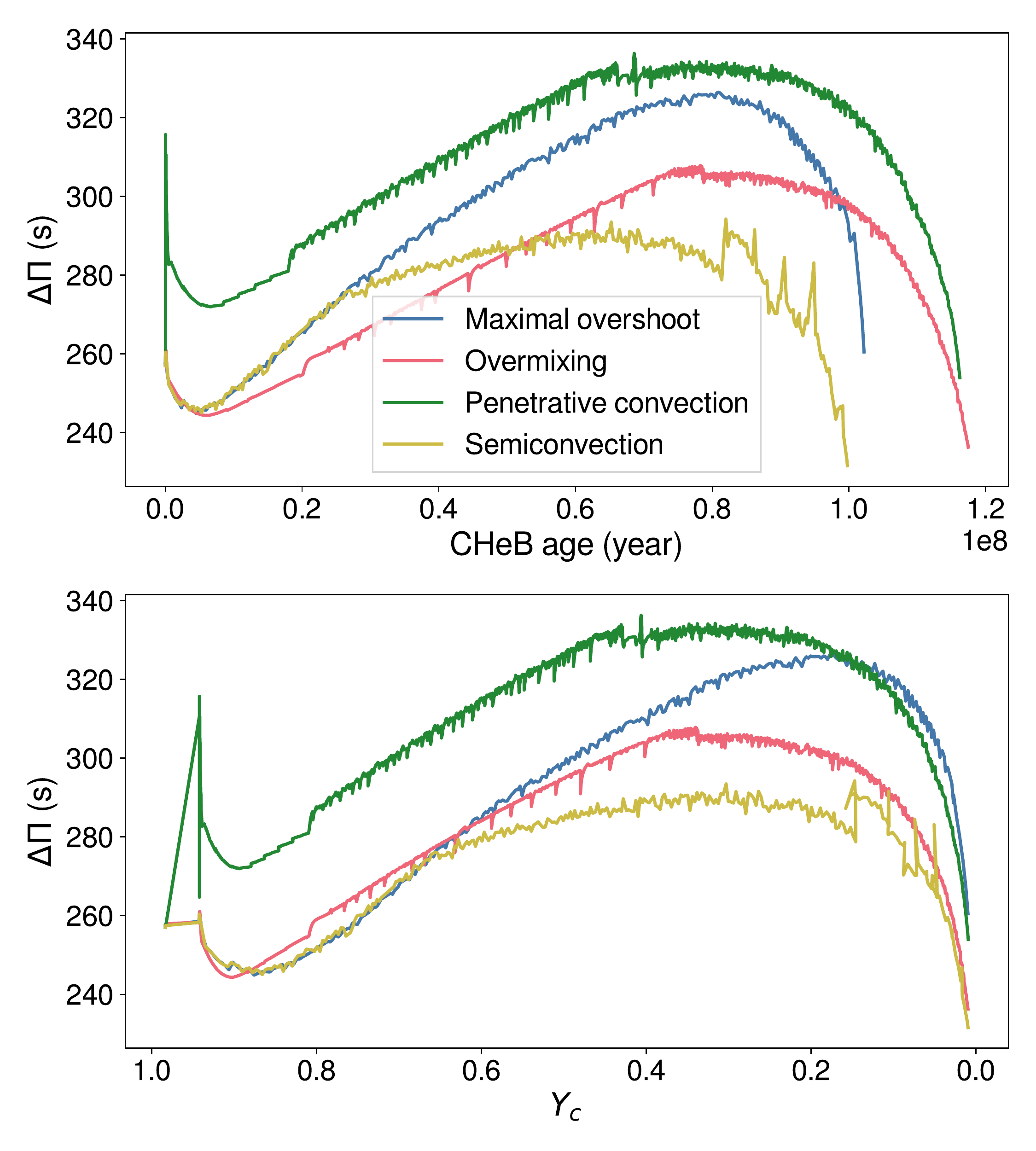}
    \caption{Evolution of $\Delta \Pi$ of a $1\,\smass$ model of solar metallicity for different core-boundary mixing prescriptions. The evolution is indicated with age (upper panel) and central helium abundance $Y_c$ (lower panel).}
    \label{evol_dp_mixing}
\end{figure}

Fig.~\ref{evol_dp_mixing} shows the evolution of the asymptotic period spacing
$\Delta \Pi$ of $1.0\,\smass$ models with solar metallicity and different core
boundary mixing prescriptions. We use $\ov = 1.0$ for overmixing and penetrative
convection. As expected, every prescription yields significantly different
ranges of $\Delta \Pi$. Notably, the maximum value of period spacing for models
with semiconvection is approximately 290s, which is significantly lower than the
highest observational values of \cite{Vrard2016}. Both maximal overshoot and
overmixing schemes show similar behavior, with a continuous growth of the core.
Finally, the penetrative convection has a behavior similar to the overmixing
scheme, but approximately $30\,\mathrm{s}$ higher.

\section{Effect of nuclear reaction rates on the seismic observables}
\label{effect_nuclear_reaction}

\subsection{Nuclear reactions in RC stars}
\begin{table}
    \caption{Nuclear rate of the $\reac$ reaction, at $10^8\,\mathrm{K}$, from
    different literature sources.}
    \label{rate_carbon_alpha}
    \begin{tabular}{l c}
        \hline \hline
        Reference & Rate value
        ($\mathrm{cm}^3\,\mathrm{s}^{-1}\,\mathrm{mol}^{-1}$) \\
        \hline
        \rule{0pt}{10pt}
        \cite{Angulo1999} &   $1.81^{+0.74}_{-0.75} \times 10^{-20} $  \\
        \rule{0pt}{10pt}
        \cite{Kunz2002} & $1.29^{+0.44}_{-0.39} \times 10^{-20} $ \\
        \rule{0pt}{10pt}
        \cite{Xu2013} &     $1.15^{+0.21}_{-0.20} \times 10^{-20} $   \\
        \rule{0pt}{10pt}
        \cite{deBoer2017} &   $1.20^{+0.20}_{-0.20} \times 10^{-20} $ \\
        \rule{0pt}{10pt}
        \cite{Shen2023} & $1.42 \times 10^{-20} $\\
        \hline
    \end{tabular}
\end{table}
Helium burning mainly happens through two reactions. The first is the 3$\alpha$
process, which occurs in two steps: $\heq + \heq \rightarrow \be$ and $\be +
\heq \rightarrow \carbon + \gamma$. This reaction dominates during the early
part of the helium burning phase. The second reaction is the $\reac$ reaction
($\carbon + \heq \rightarrow \oxygen + \gamma$) which dominates at the end of
the helium burning phase. 

The formal uncertainties on the 3$\alpha$ reaction are relatively low, being of
the order of 10\% in the NACRE compilation \citep{Angulo1999}. For the $\reac$
reaction, Table~\ref{rate_carbon_alpha} summarizes the rate at 0.1 GK found in
the literature. Notably, the work by \cite{Kunz2002} considerably lowered the
nominal rate compared to the NACRE compilation. One can also note the recent
results of \cite{Shen2023}, who found a rate significantly higher than the
values of \cite{Xu2013} and \cite{deBoer2017}. The value of the $\reac$ rate is,
therefore, still uncertain.

In this work, the standard rate for the 3$\alpha$ and  $\reac$ reactions  are
assumed to be the rates from \cite{Fynbo2005} and \cite{Xu2013}, respectively.
The models described in this section are $1\,\smass$ models with solar
metallicity, and computed with the maximal overshoot scheme.

\subsection{Effect of the nuclear reactions rates on the period spacing}

\subsubsection{3$\alpha$ Reaction}
\label{delta_p_3alpha}

\begin{figure}
    \centering
    \includegraphics[width=0.45\textwidth]{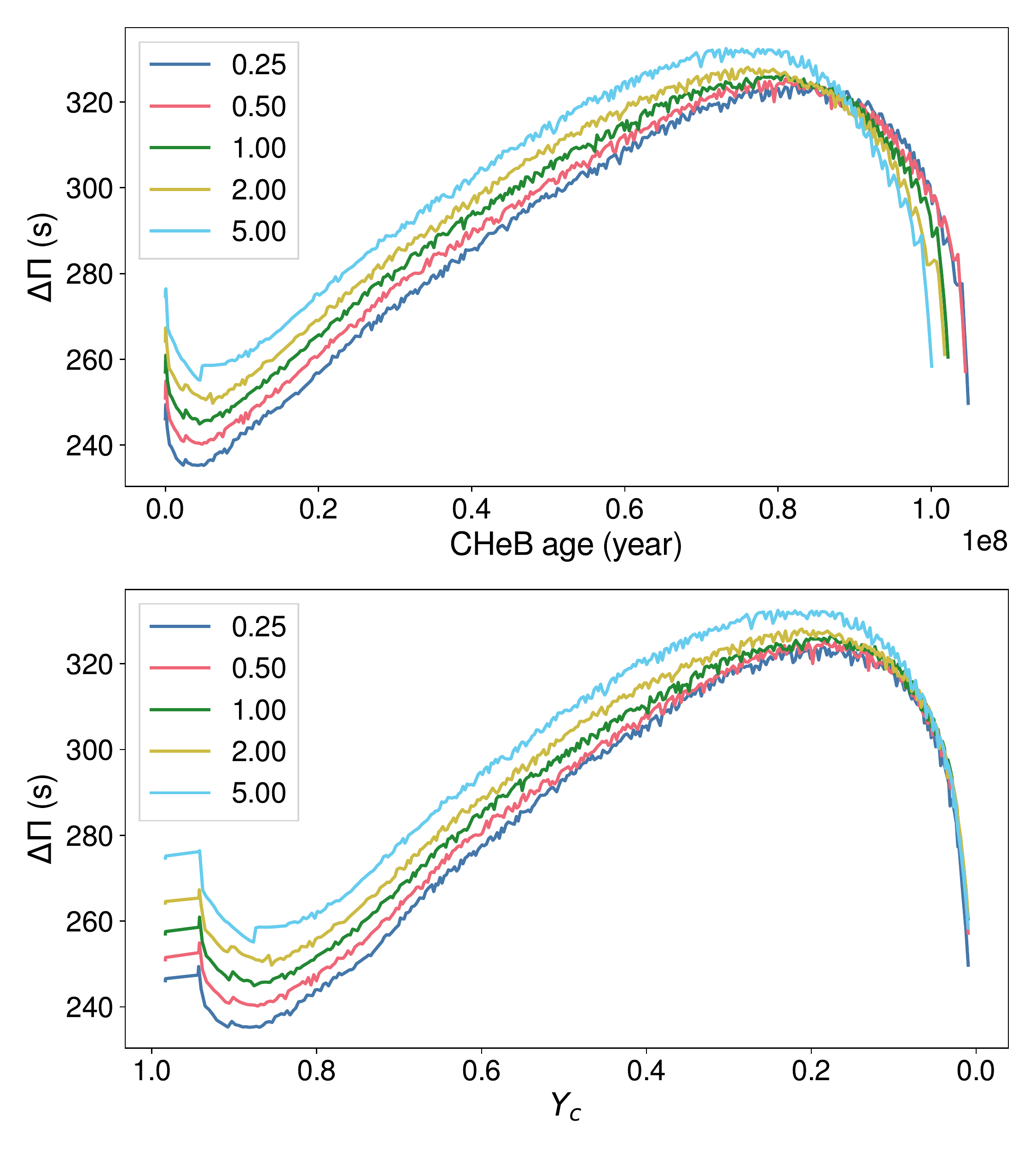}
    \caption{Evolution of $\Delta \Pi$ during the CHeB phase, for different 3$\alpha$ reaction rates.}
    \label{delta_pg_3alpha}
\end{figure}

In Fig.~\ref{delta_pg_3alpha} we show the evolution of $\Delta \Pi$ during the
core-helium burning phase for different factors by which the standard 3$\alpha$
reaction rate was changed. One can see that increasing the rate of the 3$\alpha$
process yields models with higher asymptotic period spacing during the whole
helium burning phase. For instance, multiplying the rate by 2 increases the
period spacing by approximately $4\,\mathrm{s}$, which is consistent
with the results of \cite{Constantino2015}. 

This effect on the period spacing is due to two competing processes: 1) Higher
$3\alpha$ rate allows the He-flash to start at lower temperature, i.e., earlier
in the RGB evolution. The inert He-core is therefore less massive, and so is the
convective core during the CHeB phase. This decreases the asymptotic period
spacing. 2) Because of the higher rate, the same luminosity can be produced in a
lower density environment. This lowers the local gravity within the star and
thus the \brunt frequency, eventually yielding a higher period spacing. The
amplitude of the second process is slightly larger than the first one, which
explains the fact that period spacing is higher with an increase 3$\alpha$ rate.

\subsubsection{$\calpha$ Reaction}

\begin{figure}
    \centering
    \includegraphics[width=0.45\textwidth]{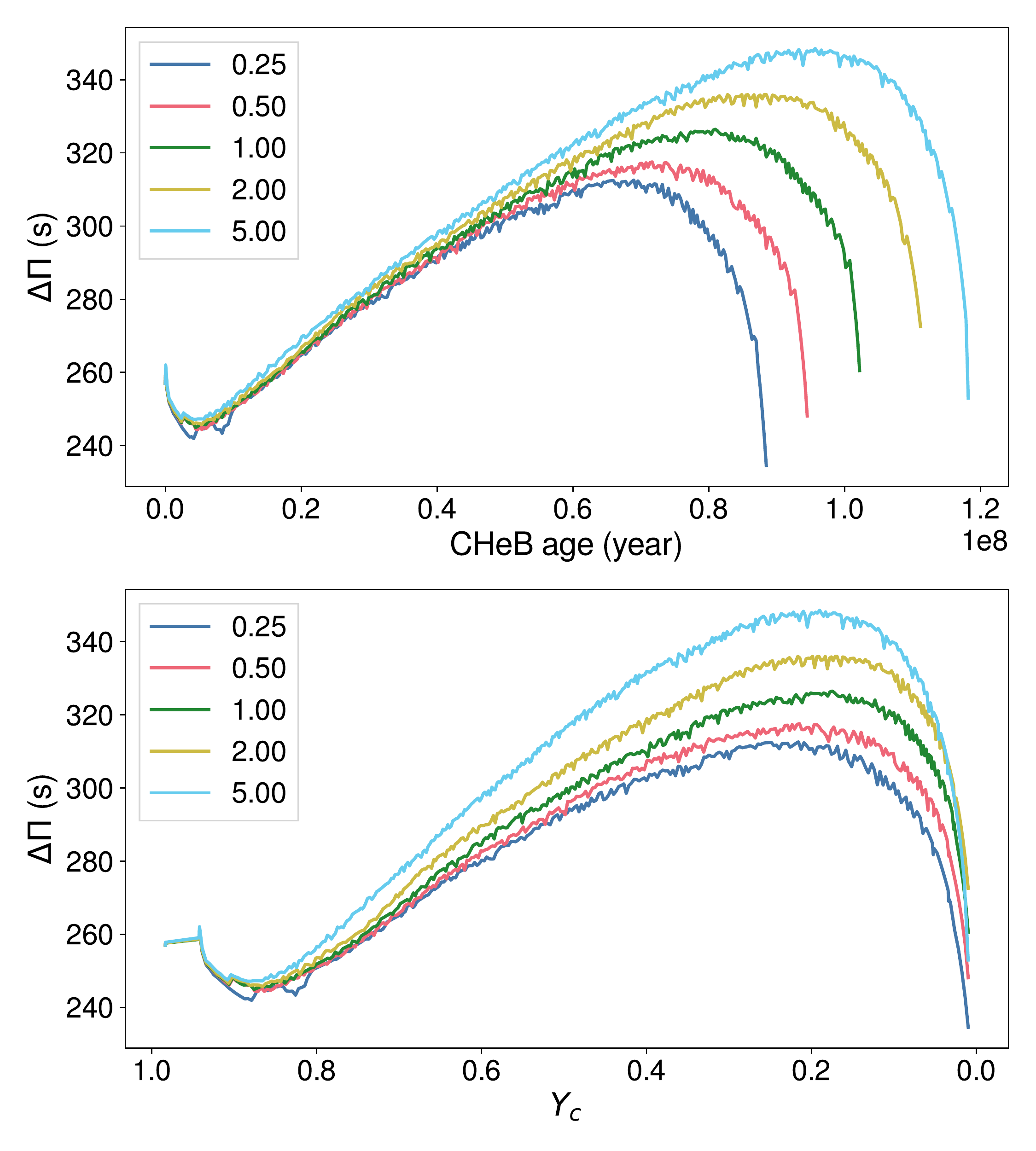}
    \caption{Same as Fig.~\ref{delta_pg_3alpha}, now for different $\reac$
    reaction rates.}
    \label{delta_pg_calpha}
\end{figure}

In Fig.~\ref{delta_pg_calpha} we show the evolution of $\Delta \Pi$ for
different rates of the $\reac$ reaction. We find that at first, for a CHeB age
below $0.4\times 10^8$ years, the period spacings of the different models are
identical. This is because initially the $\reac$ produces a negligible part of
the total luminosity of the star. The main effect of increasing the rate of
$\Delta \Pi$ is to extend the CHeB phase. Indeed, models with a rate that is
multiplied by 5 have a CHeB duration approximately 20\% longer.

The core expands during core helium burning, which increases continuously the
period spacing. Therefore, by extending the CHeB phase, the models reach higher
$\Delta \Pi$ values. Thus, models with a rate multiplied by 2 cover a period
spacing range of $89\,\mathrm{s}$, that is $10\,\mathrm{s}$ larger than the
models with a standard rate. We note that \cite{Constantino2015} found
the same order of magnitude, but surprisingly did not find that increasing the
$\reac$ rate lengthens the CHeB phase, contrary to results presented in earlier
literature \citep[e.g.,][]{Salaris2005}.

\begin{figure*}
    \centering
    \includegraphics[width=0.49\textwidth]{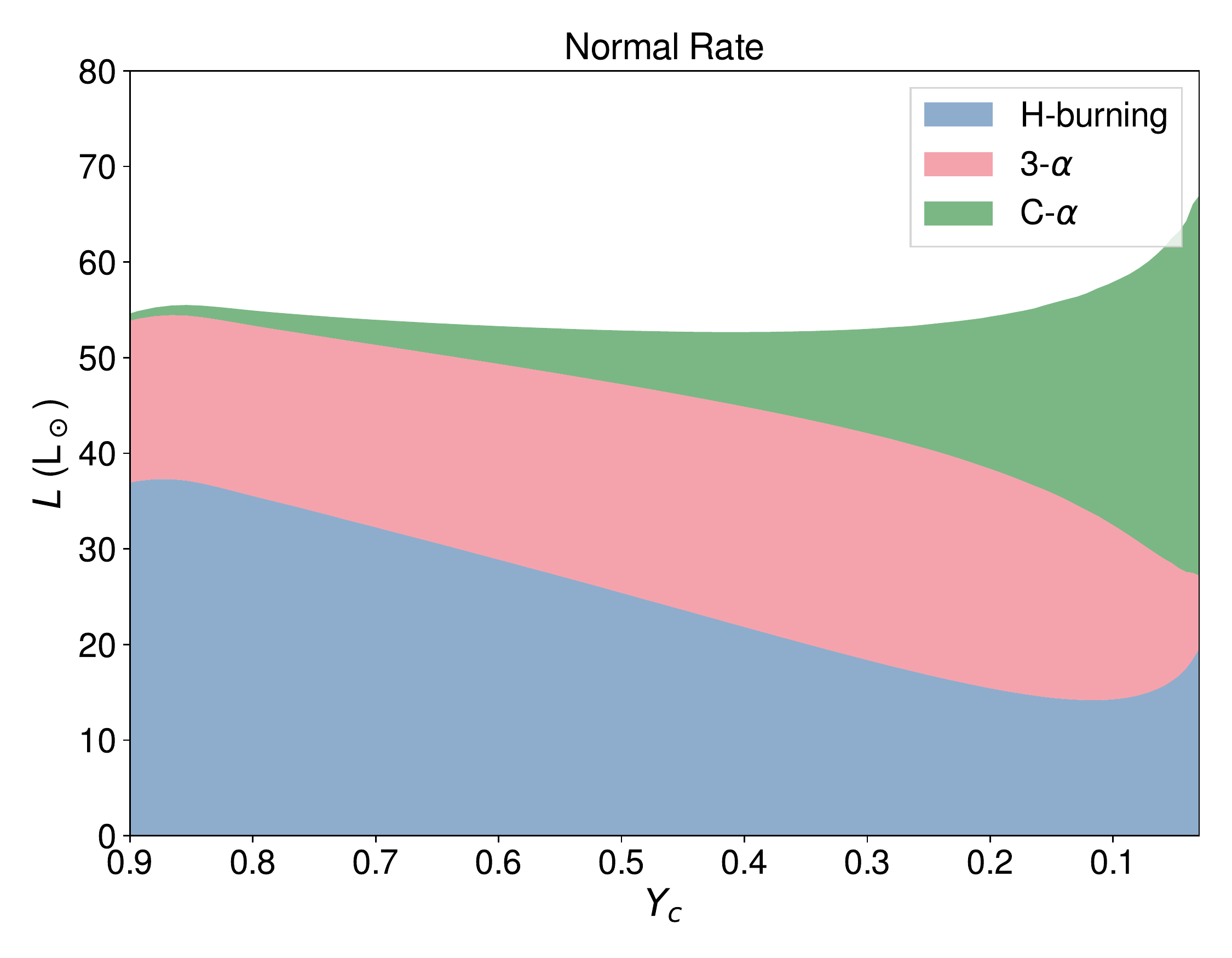}
    \includegraphics[width=0.49\textwidth]{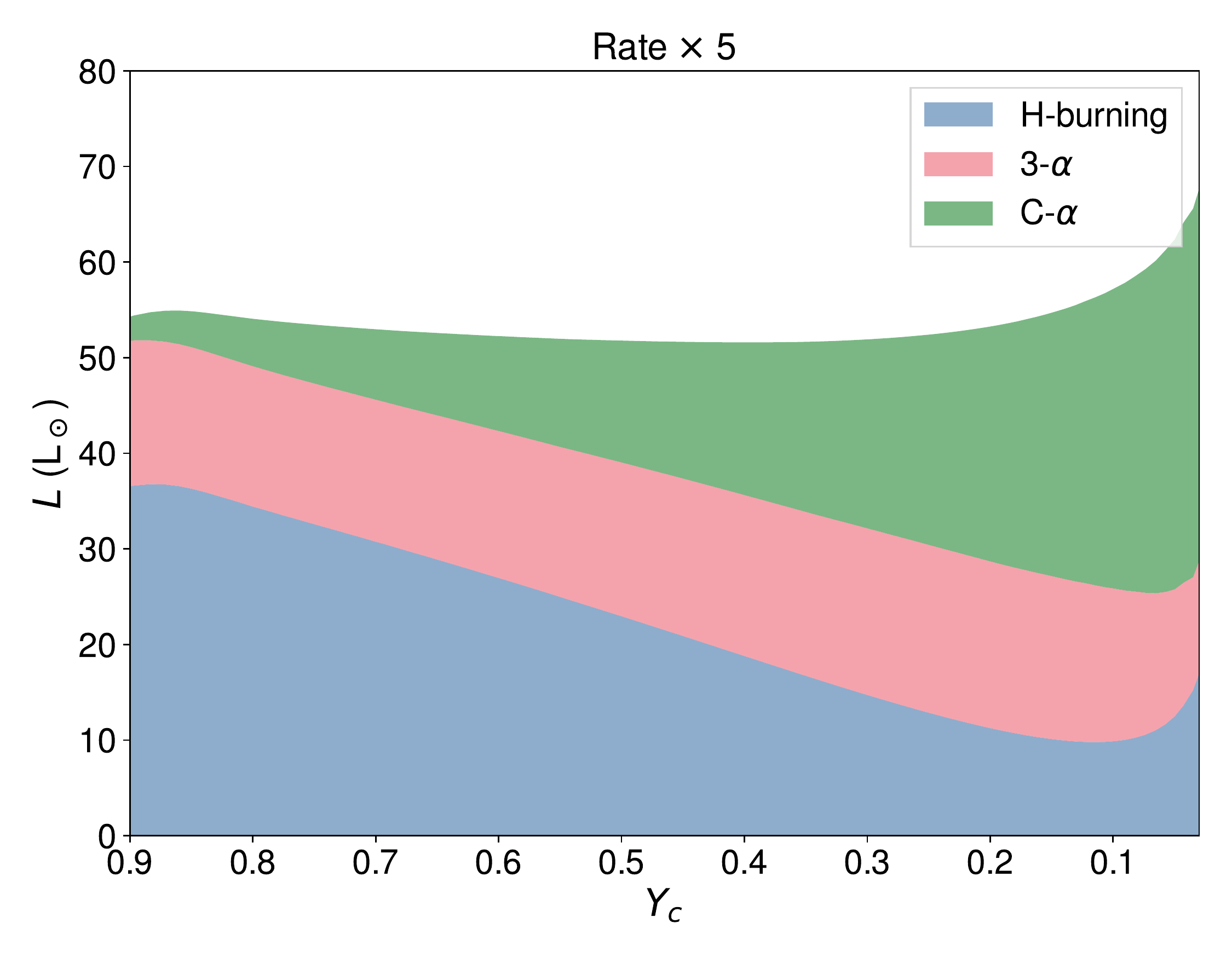}
    \caption{Evolution of the distribution of the total luminosity of a $1\,\smass$ CHeB star between the different nuclear reactions, for a star with a normal rate and a rate multiplied by 5.}
    \label{luminosity_share}
\end{figure*}

To explain the increase of the duration of the CHeB phase, we show in
Fig.~\ref{luminosity_share} the share of the luminosity production between the
different reactions, for a model with a standard rate (left) and with a rate
multiplied by 5 (right). The total luminosities of the two models are very
similar, but the distribution of energy production between the reactions differ.
Indeed, at a given $Y_c$, $\reac$ produces a more significant fraction of the
energy when its rate is multiplied by 5. Yet, one $\reac$ reaction consumes one
nucleus of helium while the 3$\alpha$ reaction consumes three, for an
approximately equal amount of energy production. Consequently, the helium
consumption will be lowered if $\reac$ dominates at a given total luminosity.
This results in an increase of the duration of the core-helium burning phase,
which in turn increases the range of $\Delta \Pi$ during the CHeB phase. We note
that similar results and explanations have recently been found by
\cite{Tognini2023}.

\section{Effect of composition}
\label{effect_other_parameters}

\begin{figure}
    \centering
    \includegraphics[width=0.49\textwidth]{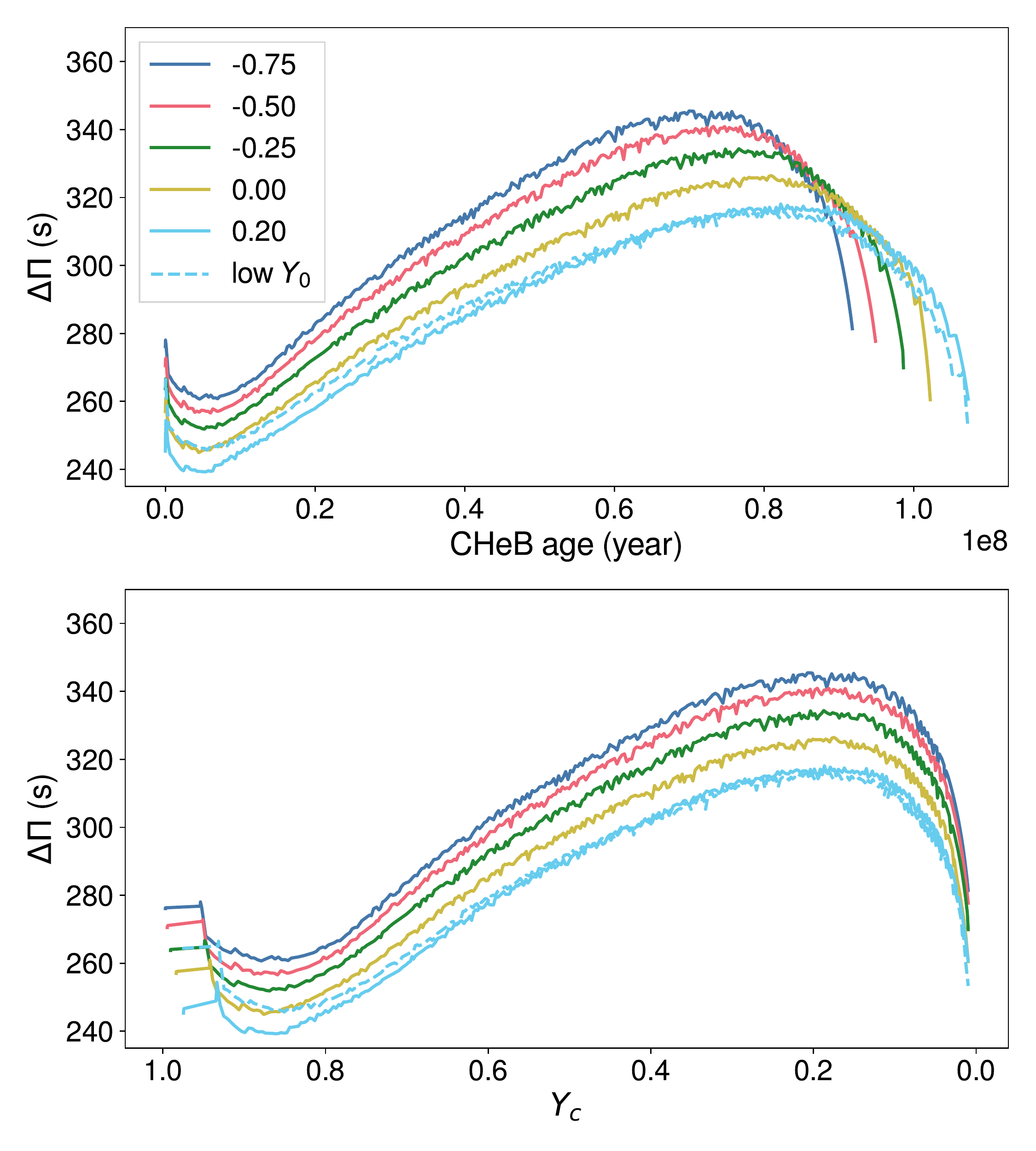}
    \caption{Evolution of $\Delta \Pi$ for different metallicities, at $1.0\,\smass$. A model with $Y_0 = 0.25$ and high metallicity is also shown.}
    \label{evol_delta_pi_metal}
\end{figure}

Metallicity has a significant impact on the period spacing during the CHeB
phase. Indeed, a metal-poor star ignites helium at a more massive core, which in
turn affects the size of the convective helium-burning core \citep[see
e.g.,][]{Sweigart1976}. Consequently, we expect a higher period spacing for
metal-poor models, which is verified in our models, as shown in
Fig.~\ref{evol_delta_pi_metal}. In addition, decreasing the metallicity shortens
the CHeB duration. Figure~\ref{evol_delta_pi_metal} shows the evolution of the
period spacing for $1\,\smass$ models, computed with a maximal overshoot scheme
at different metallicities. We found that stars with a metallicity of
$0.2\,\mathrm{dex}$ have a period spacing roughly 25\,s lower than a model with
a metallicity of $-0.5\,\mathrm{dex}$, during the whole CHeB phase.
These results are consistent with the findings of \cite{Constantino2017} and
\cite{Bossini2017}.

\medskip

In contrast with the metallicity, the models show that the initial helium $Y_0$
has a small impact on the period spacing, except at the very beginning of the
core helium burning phase. In Fig.~\ref{evol_delta_pi_metal}, we illustrate this
with a model computed with $0.2\,\mathrm{dex}$ metallicity and $Y_0 = 0.25$,
rather than the original $Y_0 = 0.29$ metallicity. 

\section{Combined effect of mixing, nuclear reaction rates and metallicity}
\label{section_combined}

\label{subsec_combined_effect}
\begin{figure*}
    \centering
    \includegraphics[width=\textwidth]{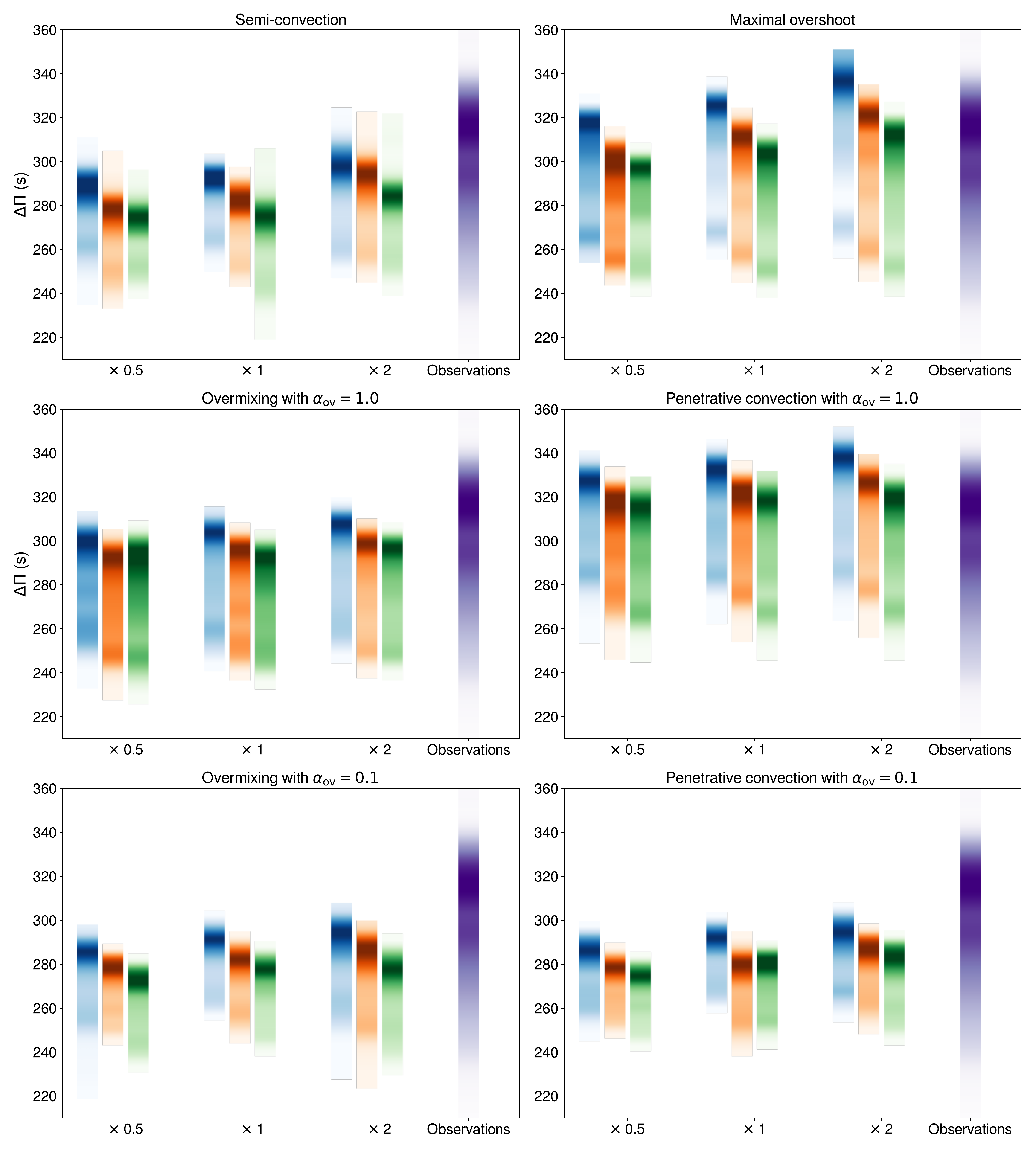} 
    \caption{$\Delta \Pi$ range for models computed with different mixing
    prescription, $\reac$ nuclear reaction rates and metallicities. Models with
    metallicities of -0.5, 0 and 0.2 dex are shown in shades of blue, orange and
    green, respectively. The opacity of the bar scales with the time that the
    model spends at that given value of period spacing. The observations from
    \cite{Vrard2016} are shown in purple, with the opacity being related to the
    number of stars observed at that value of $\Delta \Pi$. The models with
    semiconvection, overmixing and penetrative convection are computed with the
    CPM scheme.}
    \label{result_fig}
\end{figure*}

In this section, we look at the combined effect of the core boundary mixing,
nuclear reaction rates of $\reac$ and metallicity on the period spacing.
Figure~\ref{result_fig} shows the span of period spacing for models with
different mixing, metallicities and $\reac$ nuclear reaction rates. We choose
metallicity values of -0.5 and 0.2\,dex, which encompass the metallicities of
the stars in the \emph{Kepler} field. The values for the rates are half, double,
and equal to the standard rate. 

From this figure, we find that, on the one hand, the penetrative convection
scenario, when computed with $\ov = 1.0$, can reproduce the highest observed
values of period spacing, but not the lowest. On the other hand, the models
computed with penetrative convection and $\ov = 0.1$ can reproduce the lower
values but not the higher values. Thus, no value of $\ov$ seems to allow the
penetrative convection to be compatible with the observations.

The asymptotic period spacing of the models computed with semiconvection and
overmixing (for both $\ov = 1.0$ and $0.1$) cannot reproduce the full range of
observed values, especially the highest values. However, maximal overshoot
yields scenarios which are compatible with the observations. 

We note that the effect of increasing the rate is higher for models
computed with maximal overshoot (increase of the maximal value of period spacing
of $10\,\mathrm{s}$) compared to models computed with a semiconvective region
(increase of $7\,\mathrm{s}$). This effect does also depend, but only slightly,
on the metallicity, with a stronger effect for models with lower metallicity
(the increase is $2\,\mathrm{s}$ larger for $-0.5\,\mathrm{dex}$).

Finally, it seems difficult to disentangle the different reaction rates, due to
the uncertainties on the core boundary mixing and the rather strong effect of
metallicity.  

\section{Discussion}
\label{section_discussion}
\subsection{Mode trapping}
\label{trapping}

In models with overmixing or semiconvection, the PM region is part of the g-mode
cavity and its outer boundary is delimited by a strong chemical discontinuity.
For those reasons, as noted by \cite{Constantino2015}, some modes may be trapped
within the PM region while the other modes are trapped in the rest of the g-mode
cavity. Consequently, their frequencies behave as if the inner boundary of their
cavity is the outer boundary of the PM region. For this reason, the observed
period spacing, obtained with frequencies, may be significantly higher than the
asymptotic period spacing as defined in Eq.\ref{delta_pi_asymp}
\citep{Constantino2015,Pincon2022}.

We did not explicitly account for this process in our work. However, the period
spacing yielded by the maximal overshoot scheme is close to the observed period
spacing that would result from mode trapping in a semiconvective region. This is
due to the fact that the extent of the maximal overshoot core is close to the
extent of the core plus the semiconvection region. We can therefore consider
that the results coming from the maximal overshoot scheme are, at least for
$\Delta \Pi$, equivalent to the trapped mode scenario in the case of
semiconvection.

A thorough study of the mode trapping phenomenon in RC stars is beyond the scope
of this paper. We note nevertheless that we observed such trapping in a subset
of our models with semiconvection and overmixing when computing frequencies
(with, e.g., GYRE \citealt{Townsend2013}). The development of trapped modes
seems to depend, though, on the physics of the models, especially the inclusion
of microscopic diffusion.

\subsection{Effect of the core boundary scheme and $\ov$ in the overshoot models}
\label{section_effect_aov}
\begin{figure}
    \centering
    \includegraphics[width=0.45\textwidth]{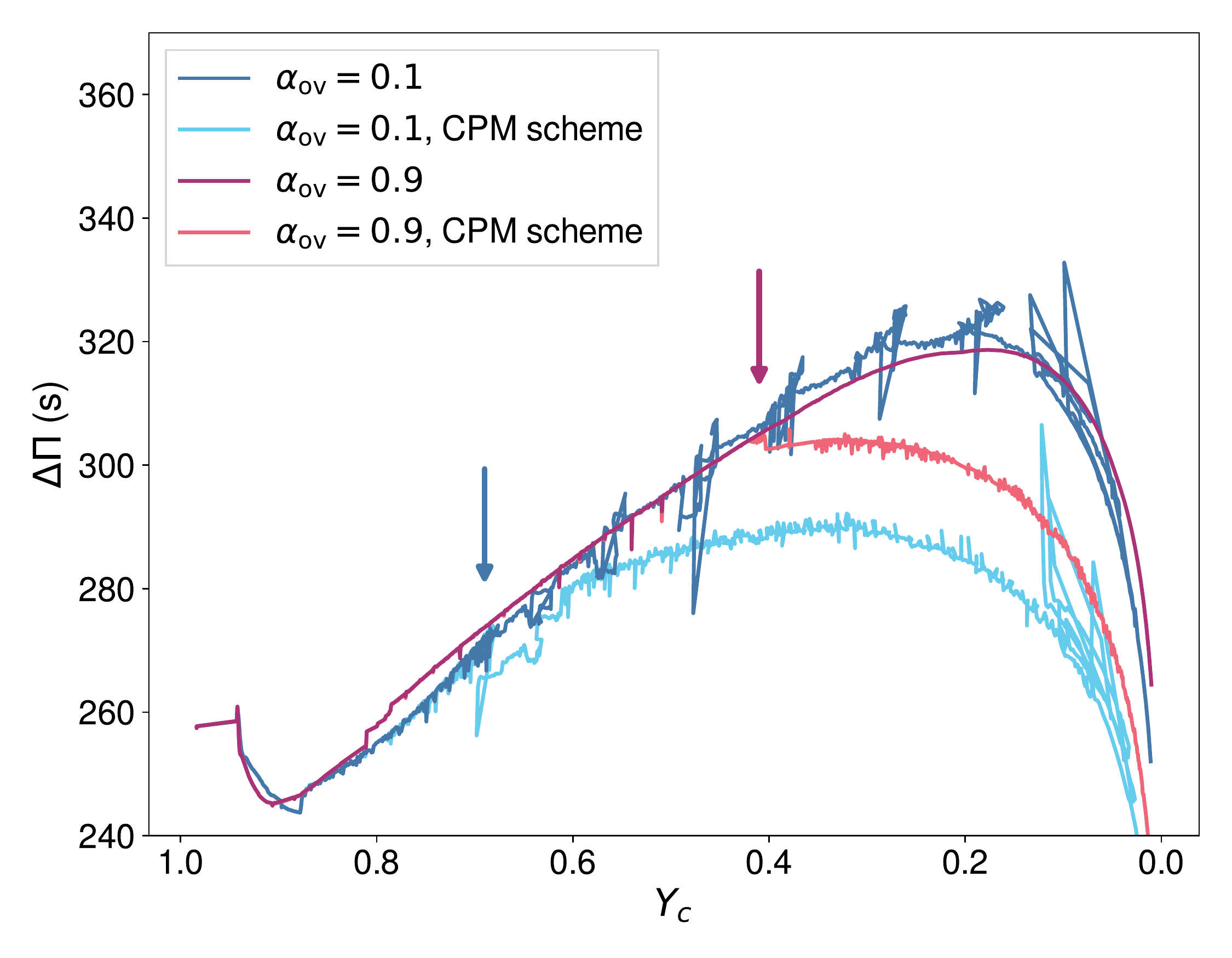}
    \caption{Evolution of the asymptotic period spacing for overmixing models
    with $\ov = 0.1$ (blue) and $\ov = 0.9$ (pink), computed with the default
    core boundary scheme (solid) and the CPM scheme (dashed) (see text for
    details). The arrows indicate the moment in the evolution from which a
    semiconvective region appears. Note the strong episodes of core breathing
    pulses (CBP) in the low overshoot models, for $Y_c < 0.15$.}
    \label{alpha_ov_effect}
\end{figure}

\begin{figure}
    \centering
    \includegraphics[width=0.45\textwidth]{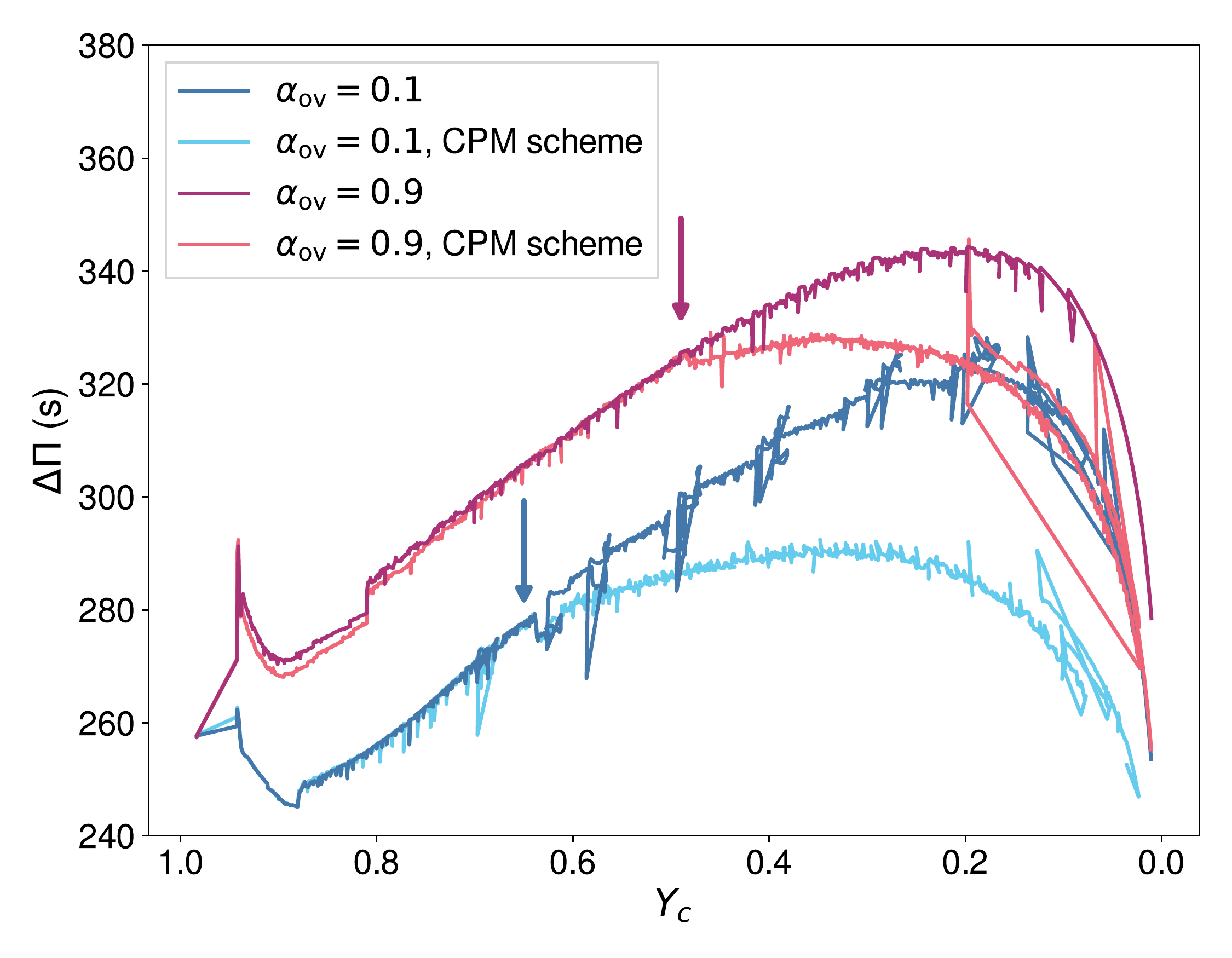}
    \caption{Same as Fig.~\ref{alpha_ov_effect}, now with penetrative convection
    ($\nabla = \gradad$ in the overshoot region.) Similarly to
    Fig.~\ref{alpha_ov_effect}, several CBP events are present.}
    \label{alpha_pc_effect}
\end{figure}
In this section, we discuss the different ways to handle the boundary of the
convective region with overshoot, as well as the effect of $\ov$.

The burning of helium in the core increases the opacity, which in turn increases
the radiative gradient. Thus, the local maximum of $\gradrad$ situated at the
outer overshoot boundary (see panel a of Fig.~\ref{mixing_properties}) becomes
at some point in the evolution larger than $\gradad$, which makes the outer part
of the overshoot region convective. Such convective shell is stable if we use
the basic convective boundary scheme, which can be questioned (see Sect.~7 of
\citealt{gabriel14}). However, if we use the convective premixing scheme (CPM),
as done in this work, the convective shell becomes semiconvective, which impacts
the \brunt frequency profile and therefore the period spacing.
Fig.~\ref{alpha_ov_effect} and \ref{alpha_pc_effect} illustrates this effect for
overmixing and penetrative convection, respectively. Notably, we can see that
using the CPM lowers the period spacing once the semiconvective region appears.
Consequently, our models have period spacings with maximal values that
are lower than the ones computed without a semiconvective region, such as in
\cite{Bossini2015,Bossini2017}.

Another effect of the CPM scheme is to make the evolution of the core boundary
smoother at low overshoot. Indeed, as noted in \cite{Straniero2003}, if a
convective shell is present in the overshoot region, it intermittently merges
with the convective core which suddenly expends it. Then, there is a strong
helium intake and the core size reduces. Such erratic core behavior is visible
in the evolution of the period spacing for models with low
overshoot\footnote{The reason such erratic evolution of $\Delta \Pi$ does not
happen at high $\ov$ is due to the fact that, as the overshoot region is larger,
the convective shell is more distant from the core and thus never merges.}, as
we can see in Fig.~\ref{alpha_ov_effect} and \ref{alpha_pc_effect}, while the
evolution of period spacing when computed with CPM appears to be smoother.

Regarding the effect of the $\ov$ parameter on the period spacing, it differs
between the overmixing and penetrative convection cases. In the first case, it
does not impact $\Delta \Pi$ much due to the fact that the overshoot region is
part of the g-mode cavity, as one can see in panel c) of
Fig.~\ref{mixing_properties}. Therefore, modifying the extent of the overshoot
region does not directly impact the extent of the g-mode cavity, which limits
the effect of the parameter on the period spacing. However, the smaller the
$\ov$ the parameter the sooner the semiconvective region appears (as noted by
the arrows), which decreases the maximum value of the period spacing as one can
see in Fig.~\ref{alpha_ov_effect}. 

For the penetrative convection case, the g-modes do not propagate in the
overshoot region. Thus, increasing $\ov$ increases the period spacing, as we can
see in Fig.~\ref{alpha_pc_effect}. Also, similarly to the models with
overmixing, the semiconvection region appears sooner in the evolution for models
with low $\ov$. Due to these two effects, $\ov$ has a considerable effect on
$\Delta \Pi$ for models with penetrative convection.

\section{Conclusion}
\label{section_conclusion}

In this paper, we investigated the combined effects of the 3$\alpha$ and
$\reac$ reaction rates, core boundary mixing and metallicity on the period
spacing of RC stars. These stars exhibit mixed modes whose
properties, and especially the period spacing $\Delta \Pi$, allow us to probe
the properties of the region around the core. We found, in accordance with the
literature \citep{Constantino2015,Constantino2017,Bossini2015,Bossini2017}, that
the core boundary mixing scheme is the main source of uncertainties on the
period spacing of our models, as the period spacing strongly depends on the size
of the convective core, as well as on the temperature and composition
stratification in the core boundary region. We confirm the findings of
\cite{Constantino2015} that the maximal overshoot is the only scheme able to
reproduce the lowest and highest values of $\Delta \Pi$. Additionally, we found
that, even if the scheme itself is ad-hoc, the resulting period spacings are
similar to those that result from mode trapping in a semiconvective region. We
also tested overmixing ($\nabla = \gradrad$ in the overshoot region) and
penetrative convection ($\nabla = \gradad$ in the overshoot region). We found
that, no single value of the overshoot parameter can reproduce both the highest
and lowest observed values of $\Delta \Pi$. These results differ from those of
\cite{Bossini2015,Bossini2017}. This difference is due to the use of the
convective premixing scheme in this work, which causes a semiconvective region
around the core which decreases the value of the period spacing. This
semiconvection region notably solves the issue of an unphysical core boundary,
in which $\gradrad > \gradad$ at the inner side of the boundary.

We also investigated the effect of changing the rates of the 3$\alpha$
and $\reac$ reactions. On the one hand, regarding the 3$\alpha$ reaction, a
higher rate causes an earlier helium flash and a lower central density during
the helium-burning phase. These effects decrease and increase the period
spacing, respectively, such that the resulting effect is negligible. On the
other hand, the effect of increasing the rate of the $\reac$ reaction is,
mainly, to increase the duration of the CHeB phase. This allows the cores to
grow to a larger size, which eventually increases the maximum value of the
period spacing. We find that doubling the rate increases the maximum period
spacing value by $10\,\mathrm{s}$ for a model computed with maximal overshoot,
and $7\,\mathrm{s}$ for a model computed with the convective premixing scheme
without overshoot. The effect of increasing the rate is slightly higher for
models with lower metallicity, but with a negligible amplitude ($2\,\mathrm{s}$
higher for a -0.5 dex model).

Mainly due to the uncertainties on the mixing\footnote{We note the recent work
from \cite{Blouin2023}, which through 3D simulations, explore the stability of
the different core boundary mixing schemes used in this work, and find that a
semiconvective region is quickly erased by overshooting mixing through
overshooting from the convective core. }, it is for the moment not (yet)
possible to perform an inverse analysis, i.e., constrain the $\reac$ rate using
the period spacing of the RC stars. However, with better constraints on
the core boundary mixing, as well as more data with the upcoming PLATO mission
\citep{plato}, RC stars seismology could provide astrophysical
constraints on the $\reac$ reaction rate.

\begin{acknowledgements}
	We thank the anonymous referee for comments that improved the clarity of
this paper. AN and SH acknowledge funding from the ERC Consolidator Grant
DipolarSound (grant agreement \#10s1000296). SB acknowledges NSF grant
AST-2205026.
\end{acknowledgements}

\bibliographystyle{aa} 
\bibliography{biblio} 
\end{document}